\documentstyle[12pt]{article}
\renewcommand{\baselinestretch}{1.}

\begin{document}
\parskip=5pt plus 1pt minus 1pt

\begin{flushright}
{\bf hep-ph/9909304} \\
{\bf LMU-99-09}\\
September 1999
\end{flushright}

\vspace{0.2cm}

\begin{center}
{\Large\bf Maximal Neutrino Mixing and Maximal $CP$ Violation}
\end{center}

\vspace{0.5cm}
\begin{center}
{\bf Harald Fritzsch} 
\footnote{Electronic address: bm@hep.physik.uni-muenchen.de}
~ and ~ {\bf Zhi-zhong Xing}
\footnote{Electronic address: xing@hep.physik.uni-muenchen.de}
\\
{\sl Sektion Physik, Universit$\sl\ddot{a}$t M$\sl\ddot{u}$nchen,
Theresienstrasse 37A, 80333 M$\sl\ddot{u}$nchen, Germany}
\end{center}

\vspace{3cm}
\begin{abstract}
We propose a phenomenological model of lepton mixing and $CP$ 
violation based on 
the flavor democracy of charge leptons and the
mass degeneracy of neutrinos. A 
nearly bi-maximal flavor mixing pattern, which is favored by current
data on atmospheric and solar neutrino oscillations, emerges naturally
from this model after explicit symmetry breaking.
The rephasing-invariant strength of $CP$ or $T$ violation 
can be as large as one percent, leading to significant
probability asymmetries between $\nu_\mu \rightarrow \nu_e$
and $\bar{\nu}_\mu \rightarrow \bar{\nu}_e$ (or 
$\nu_e \rightarrow \nu_\mu$) transitions in the long-baseline
neutrino experiments.  
\end{abstract}

\newpage

The recent observation of atmospheric and solar neutrino
anomalies, in particular by the Super-Kamiokande experiment \cite{SK},
has provided a strong indication that neutrinos are massive and 
lepton flavors are mixed. As there exist at least three 
different lepton families, the flavor mixing matrix may 
in general contain non-trivial complex phase terms. Hence $CP$
or $T$ violation is naturally expected in the lepton sector.

A violation of $CP$ invariance in the quark sector can result in
a variety of observable effects in hadronic weak decays. 
Similarly $CP$ or $T$ 
violation in the lepton sector can manifest itself
in neutrino oscillations \cite{Cabibbo}.
The best (and probably the only) way to observe $CP$- or
$T$-violating effects in neutrino oscillations is to carry out
the long-baseline appearance neutrino experiments \cite{CP}.

In the scheme of three lepton families, the $3\times 3$ flavor
mixing matrix $V$ links the neutrino mass eigenstates
$(\nu_1, \nu_2, \nu_3)$ to the neutrino flavor eigenstates
$(\nu_e, \nu_\mu, \nu_\tau)$:
\begin{equation}
\left ( \matrix{
\nu_e \cr
\nu_\mu \cr
\nu_\tau \cr} \right ) \; =\; 
\left ( \matrix{
V_{e1}	& V_{e2}	& V_{e3} \cr
V_{\mu 1}	& V_{\mu 2}	& V_{\mu 3} \cr
V_{\tau 1}	& V_{\tau 2}	& V_{\tau 3} \cr} \right )
\left ( \matrix{
\nu_1 \cr
\nu_2 \cr
\nu_3 \cr} \right ) \; .
%		(1)
\end{equation}
If neutrinos are massive Dirac fermions, $V$ can be parametrized
in terms of three rotation angles and one $CP$-violating phase.
If neutrinos are Majorana fermions, however, two additional
$CP$-violating phases are in general needed to fully parametrize $V$.
The strength of $CP$ violation in neutrino oscillations, no
matter whether neutrinos are of the Dirac or Majorana type, 
depends only upon a universal parameter $\cal J$ \cite{Jarlskog}, which 
is defined through
\begin{equation}
{\rm Im} \left (V_{il}V_{jm} V^*_{im}V^*_{jl} \right )
\; =\; {\cal J} \sum_{k,n} \epsilon^{~}_{ijk} \epsilon^{~}_{lmn} 
\; .
%		(2)
\end{equation}
The asymmetry between the probabilities of two $CP$-conjugate 
neutrino transitions, due to the $CPT$
invariance and the unitarity of $V$, is uniquely given as 
\begin{eqnarray}
\Delta_{CP} & = & P(\nu_\alpha \rightarrow \nu_\beta) - P(\bar{\nu}_\alpha
\rightarrow \bar{\nu}_\beta) \; \nonumber \\
& = & -16 {\cal J} \sin F_{12} \sin F_{23} \sin F_{31} \; 
%		(3)
\end{eqnarray}
with $(\alpha, \beta) = (e,\mu)$, $(\mu, \tau)$ or $(\tau, e)$,
$F_{ij} = 1.27 \Delta m^2_{ij} L/E$ and
$\Delta m^2_{ij} = m^2_i -m^2_j$, in which $L$ is the distance
between the neutrino source and the detector
(in unit of km) and $E$ denotes the neutrino beam energy (in unit of
GeV). The $T$-violating asymmetry can be obtained in a
similar way
%%%%%%%%%%%%%%% 
\footnote{Note that an asymmetry between the probabilities 
$P(\bar{\nu}_\alpha \rightarrow \bar{\nu}_\beta)$ and 
$P(\bar{\nu}_\beta
\rightarrow \bar{\nu}_\alpha)$ signifies $T$ violation too. 
This asymmetry and that defined in Eq. (4) have the same magnitude
but opposite signs.} :
%%%%%%%%%%%%%%%
\begin{eqnarray}
\Delta_T & = & P(\nu_\alpha \rightarrow \nu_\beta)
- P(\nu_\beta \rightarrow \nu_\alpha) \; \nonumber \\
& = & -16 {\cal J} \sin F_{12} \sin F_{23} \sin F_{31} \; .
%		(4)
\end{eqnarray}
These formulas show clearly that $CP$ or $T$ violation is a feature
of all three lepton families. The relationship $\Delta_T 
= \Delta_{CP}$ is a straightforward consequence
of $CPT$ invariance. 
The observation of $\Delta_T$ might be free from the
matter effects of the earth, which is possible to fake the genuine
$CP$ asymmetry $\Delta_{CP}$ in any long-baseline neutrino
experiment. The joint measurement of $\nu_\alpha \rightarrow
\nu_\beta$ and $\nu_\beta \rightarrow \nu_\alpha$ transitions to
determine $\Delta_T$ is, however, a challenging task in practice.
Probably it could only be realized in a neutrino factory,
whereby high-quality neutrino beams can be produced with high-intensity 
muon storage rings \cite{Muon}.

Analyses of current experimental data \cite{SK,CHOOZ} 
%%%%%%%%%%%%%%%%%%%%%%%%%%%
\footnote{Throughout this work we do not take the LSND evidence
for neutrino oscillations \cite{LSND}, which was not confirmed by the
KARMEN experiment \cite{KARMEN}, into account.}
%%%%%%%%%%%%%%%%%%%%%%%%%%%
yield $\Delta m^2_{\rm sun} \ll \Delta m^2_{\rm atm}$ and
$|V_{e3}|^2 \ll 1$, implying that the atmospheric and solar
neutrino oscillations are approximately decoupled.
A reasonable interpretation of those data follows from setting
$\Delta m^2_{\rm sun} = |\Delta m^2_{12}|$ and
$\Delta m^2_{\rm atm} = |\Delta m^2_{23}| \approx |\Delta m^2_{31}|$.
In this approximation $F_{31} \approx -F_{23}$ holds.
The $CP$- and $T$-violating asymmetries can then be 
simplified as 
\begin{equation}
\Delta_{CP} \; =\; \Delta_T \; \approx \; 
16 {\cal J} \sin F_{12} \sin^2 F_{23} \; .
%		(5)
\end{equation}
Note that $\Delta_{CP}$ or $\Delta_T$ depends linearly on
the oscillating term $\sin F_{12}$, 
therefore the length of the baseline suitable for
measuring $CP$ and $T$ asymmetries should satisfy the 
condition $|\Delta m^2_{12}| \sim E/L$.
This requirement singles out 
the large-angle MSW solution, which has $\Delta m^2_{\rm sun} \sim 10^{-5}$
to $10^{-4} ~ {\rm eV}^2$ and $\sin^2 2\theta_{\rm sun} \sim 0.65$ 
to $1$ \cite{Bahcall}, among
three possible solutions to the solar neutrino problem.
The small-angle MSW solution is not favored; it does not give rise to 
a relatively large magnitude of $\cal J$, which determines 
the significance of practical $CP$- or $T$-violating signals.
The long wave-length vacuum oscillation requires $\Delta m^2_{\rm sun}
\sim 10^{-10}$ eV$^2$, too small to meet the realistic long-baseline
prerequisite.

In this paper we extend our previous hypothesis of lepton flavor 
mixing \cite{FX96}, which arises naturally from the breaking of flavor 
democracy for charged leptons and that of mass degeneracy for
neutrinos, to include $CP$ violation. It is found that the 
rephasing-invariant strength of $CP$ or $T$ violation can be as large
as one percent. The flavor mixing pattern remains 
nearly bi-maximal, thus both atmospheric and solar neutrino
oscillations can well be interpreted. The consequences of 
the model on the future long-baseline neutrino experiments will also
be discussed in some detail.

The phenomenological constraints obtained from various neutrino
oscillation experiments indicate that the mass differences in the
neutrino sector are tiny compared to those in the charged
lepton sector. One possible interpretation is that all three
neutrinos are nearly degenerate in mass. In this case one 
might expect that the flavor mixing pattern of leptons differs
qualitatively from that of quarks, where both up and down
sectors exhibit a strong hierarchical structure in their mass
spectra and the observed mixing angles are rather small. 
A number of authors have argued that the hierarchy of quark masses
and the smallness of mixing angles are related to each other,
by considering specific symmetry limits \cite{FX99}. One particular
way to proceed is to consider the limit of subnuclear democracy,
in which the mass matrices of both the up- and down-type quarks
are of rank one and have the structure
\begin{equation}
M_{\rm q} \; =\; \frac{c_{\rm q}}{3} \left ( \matrix{
1	& 1	& 1 \cr
1	& 1	& 1 \cr
1	& 1	& 1 \cr} \right )
%		(6)
\end{equation}
with q = u (up) or d (down) as well as $c_{\rm u} = m_t$
and $c_{\rm d} = m_b$. 
Small departures from the democratic
limit lead to the flavor mixing and at the same time introduce 
the masses of the second and first families. Specific symmetry
breaking schemes have been proposed in some literature
in order to calculate the flavor mixing angles in terms of
the quark mass eigenvalues (see, e.g., Ref. \cite{FX99}).

Since the charged leptons exhibit a similar hierarchical 
mass spectrum as the quarks, it would be natural to consider
the limit of subnuclear democracy for the 
$(e, \mu, \tau)$ system, i.e., the mass matrix takes the
form as Eq. (6). In the same 
limit three neutrinos are degenerate in mass. Then we have \cite{FX96}
\begin{eqnarray}
M^{(0)}_l & = & \frac{c^{~}_l}{3} \left (\matrix{
1	& 1	& 1 \cr
1	& 1	& 1 \cr
1	& 1	& 1 \cr} \right ) \; , 
\nonumber \\
M^{(0)}_\nu & = & c_\nu \left (\matrix{
1	& 0	& 0 \cr
0	& 1	& 0 \cr
0	& 0	& 1 \cr} \right ) \; ,
%		(7)
\end{eqnarray}
where $c^{~}_l =m_\tau$ and $c_\nu =m_0$ measure the 
corresponding mass scales.
If the three neutrinos are of the Majorana type,
$M^{(0)}_\nu$ could take a more general form
$M^{(0)}_\nu P_\nu$ with $P_\nu = {\rm Diag} \{ 1,
e^{i\phi_1}, e^{i\phi_2} \}$. As the Majorana phase matrix
$P_\nu$ has no effect on the flavor mixing and
$CP$-violating observables
in neutrino oscillations, it will be neglected in the subsequent discussions.
Clearly $M^{(0)}_\nu$ exhibits an
S(3) symmetry, while $M^{(0)}_l$ an
$S(3)_{\rm L} \times S(3)_{\rm R}$ symmetry.

One can transform the charged lepton mass matrix from the 
democratic basis $M^{(0)}_l$ into the hierarchical basis
\begin{equation}
M^{(\rm H)}_l \; =\; c^{~}_l \left ( \matrix{
0	& 0 	& 0 \cr
0	& 0 	& 0 \cr
0	& 0 	& 1 \cr } \right ) 
%		(8)
\end{equation}
by means of an orthogonal transformation, i.e.,
$M^{(\rm H)}_l = U M^{(0)}_l U^{\rm T}$, where
\begin{equation}
U \; =\; \left ( \matrix{
\frac{1}{\sqrt{2}}	& ~~ \frac{-1}{\sqrt{2}} ~	& 0 \cr
\frac{1}{\sqrt{6}}	& ~~ \frac{1}{\sqrt{6}} ~	& \frac{-2}{\sqrt{6}} \cr
\frac{1}{\sqrt{3}}	& ~~ \frac{1}{\sqrt{3}} ~	& \frac{1}{\sqrt{3}} \cr}
\right ) \; .
%		(9)
\end{equation}
We see $m_e = m_\mu =0$ from $M^{(\rm H)}_l$ and
$m_1 = m_2 =m_3 =m_0$ from $M^{(0)}_\nu$. Of course
there is no flavor mixing in this symmetry limit.

A simple real diagonal breaking of the flavor democracy
for $M^{(0)}_l$ and the mass degeneracy for $M^{(0)}_\nu$
may lead to instructive results for flavor mixing 
in neutrino oscillations \cite{FX96,Tanimoto}.
To accommodate $CP$ violation, however, complex perturbative
terms are required. 
Let us proceed with two different symmetry-breaking steps
in close analogy to the symmetry breaking discussed 
previously for the quark mass matrices \cite{FP90,FH94}.

First, small real perturbations to the (3,3) elements of $M^{(0)}_l$
and $M^{(0)}_\nu$ are respectively introduced:
\begin{eqnarray}
\Delta M^{(1)}_l & = & \frac{c^{~}_l}{3} \left ( \matrix{
0	& 0	& 0 \cr
0	& 0	& 0 \cr
0	& 0	& \varepsilon^{~}_l \cr } \right ) \; , 
\nonumber \\
\Delta M^{(1)}_\nu & = & c_\nu \left ( \matrix{
0	& 0	& 0 \cr
0	& 0	& 0 \cr
0	& 0	& \varepsilon_\nu \cr } \right ) \; .
%		(10)
\end{eqnarray}
In this case the charged lepton mass matrix $M^{(1)}_l =
M^{(0)}_l + \Delta M^{(1)}_l$ remains symmetric under an
$S(2)_{\rm L}\times S(2)_{\rm R}$ transformation, 
and the neutrino mass matrix
$M^{(1)}_\nu = M^{(0)}_\nu + \Delta M^{(0)}_\nu$ has
an $S(2)$ symmetry. 
The muon becomes massive (i.e., $m_\mu \approx 2|\varepsilon^{~}_l|
m_\tau /9$), and the mass eigenvalue $m_3$ is no more degenerate
with $m_1$ and $m_2$ (i.e., $|m_3 - m_0| = m_0 |\varepsilon_\nu|$). 
After the diagonalization of 
$M^{(1)}_l$ and $M^{(1)}_\nu$, one finds that the 2nd and 3rd
lepton families have a definite flavor mixing angle
$\theta$. We obtain $\tan\theta \approx -\sqrt{2} ~$ if the
small correction of ${\cal O}(m_\mu/m_\tau)$ is neglected.
Then neutrino oscillations at the atmospheric scale may arise
in $\nu_\mu \rightarrow \nu_\tau$ transitions with 
$\Delta m^2_{32} = \Delta m^2_{31}
\approx 2m_0 |\varepsilon_\nu|$. The corresponding
mixing factor $\sin^2 2\theta \approx 8/9$ is in good agreement 
with current data.

The symmetry breaking given in Eq. (10) for the charged lepton
mass matrix serves as a good illustrative example. One could 
consider a more general case, analogous to the one for 
quarks \cite{FP90}, to break the $S(3)_{\rm L}\times S(3)_{\rm R}$
symmetry of $M^{(0)}_l$ to an $S(2)_{\rm L} \times S(2)_{\rm R}$
symmetry. This would imply that $\Delta M^{(1)}_l$ takes the
form
\begin{equation}
\Delta M^{(1)}_l \; =\; \frac{c^{~}_l}{3}
\left ( \matrix{
0	& 0	& \varepsilon'_l \cr
0	& 0	& \varepsilon'_l \cr
\varepsilon'_l	& \varepsilon'_l	& \varepsilon^{~}_l \cr}
\right ) \; ,
%		(11)
\end{equation}
where $|\varepsilon^{~}_l| \ll 1$ and $|\varepsilon'_l| \ll 1$. 
In this case the leading-order results obtained above, i.e.,
$\tan\theta \approx -\sqrt{2}$ 
and $\sin^2 2\theta \approx 8/9$, remain unchanged. 

At the next step we introduce a complex symmetry breaking 
perturbation, analogous to that for quark mass matrices
discussed in Ref. \cite{Lehmann}, to the charged lepton mass 
matrix $M^{(1)}_l$:
\begin{equation}
\Delta M^{(2)}_l \; = \; \frac{c^{~}_l}{3} \left ( \matrix{
0	& ~ -i\delta_l ~	& i\delta \cr
i\delta	& ~ 0 ~		& -i\delta_l \cr
-i\delta_l	& ~ i\delta_l ~	& 0 \cr } \right ) \; .
%		(12)
\end{equation}
Transforming $M^{(2)}_l = M^{(1)}_l + \Delta M^{(2)}_l$ into
the hierarchical basis, we obtain
\begin{equation}
M^{\rm H}_l \; = \; c^{~}_l \left ( \matrix{
0	& -i\frac{1}{\sqrt{3}}\delta_l	& 0 \cr
i\frac{1}{\sqrt{3}}\delta_l	& \frac{2}{9}\varepsilon^{~}_l
& -\frac{\sqrt{2}}{9}\varepsilon^{~}_l \cr
0	& -\frac{\sqrt{2}}{9}\varepsilon^{~}_l
& 1 + \frac{1}{9}\varepsilon^{~}_l \cr} 
\right ) \; .
%		(13)
\end{equation}
Note that $M^{\rm H}_l$, just like a variety of realistic
quark mass matrices \cite{FX99}, has texture zeros in the
(1,1), (1,3) and (3,1) positions. 
The phases of its (1,2) and (2,1) elements are $\mp 90^{\circ}$, 
which could lead to maximal $CP$ violation if the neutrino mass 
matrix is essentially real. 
For the latter we consider a small perturbation, analogous to
that in Eq. (10), to break the remaining mass degeneracy of
$M^{(1)}_\nu$:
\begin{equation}
\Delta M^{(2)}_\nu \; = \; c_\nu \left ( \matrix{
-\delta_\nu	& ~ 0 	&  0 \cr
0	& ~ \delta_\nu 	&  0 \cr
0	&  ~ 0 	&  0 \cr } \right ) \; .
%		(14)
\end{equation}
From $\Delta M^{(2)}_l$ and $\Delta M^{(2)}_\nu$ 
we obtain $m_e \approx |\delta_l|^2 m^2_\tau /(27 m_\mu)$ 
and $m_{1,2} = m_0 (1 \mp \delta_\nu)$, respectively. 
The simultaneous diagonalization of 
$M^{(2)}_l = M^{(1)}_l + \Delta M^{(2)}_l$ and 
$M^{(2)}_\nu = M^{(1)}_\nu + \Delta M^{(2)}_\nu$ 
leads to a full $3\times 3$
flavor mixing matrix, which links neutrino mass eigenstates 
$(\nu_1, \nu_2, \nu_3)$ to neutrino flavor eigenstates
$(\nu_e, \nu_\mu, \nu_\tau)$ in the following manner:
\begin{equation}
V \; =\; U \; + \; i ~ \xi^{~}_V \sqrt{\frac{m_e}{m_\mu}} 
\; + \; \zeta^{~}_V \frac{m_\mu}{m_\tau} \; ,
%		(15)
\end{equation}
where $U$ has been given in Eq. (9), and
\begin{eqnarray}
\xi^{~}_V & = &  \left ( \matrix{
\frac{1}{\sqrt{6}}	& ~ \frac{1}{\sqrt{6}} ~	& \frac{-2}{\sqrt{6}} \cr
\frac{1}{\sqrt{2}}	& ~ \frac{-1}{\sqrt{2}} ~	& 0 \cr
0	& ~ 0 ~	& 0 \cr} \right ) \; ,
\nonumber \\
\zeta^{~}_V & = & \left ( \matrix{
0	& 0	& 0 \cr
\frac{1}{\sqrt{6}}	& \frac{1}{\sqrt{6}}	& \frac{1}{\sqrt{6}} \cr
\frac{-1}{\sqrt{12}}	& \frac{-1}{\sqrt{12}}	& \frac{1}{\sqrt{3}} \cr}
\right ) \; .
%		(16)
\end{eqnarray}
In comparison with the result of Ref. \cite{FX96},
the new feature of this lepton mixing scenario is that 
the term multiplying $\xi^{~}_V$ becomes imaginary. Therefore $CP$ or $T$
violation has been incorporated. 

The complex symmetry breaking perturbation given in Eq. (12) is
certainly not the only one which can be considered for $M^{(1)}_l$.
Below we list a number of other interesting possibilities, i.e., 
the hermitian perturbations
\begin{eqnarray}
\Delta \tilde{M}^{(2)}_l & = & \frac{c^{~}_l}{3} \left ( \matrix{
~ 0 ~	& -i\delta_l ~	& ~ 0 ~~ \cr
~ i\delta_l ~	& ~ 0 ~	& ~ 0 ~~ \cr
~ 0 ~	& ~ 0 ~	& ~ 0 ~~ \cr } \right ) \; , 
\nonumber \\
\Delta \hat{M}^{(2)}_l & = & \frac{c^{~}_l}{3} \left ( \matrix{
~ 0	& ~ 0 ~	& i\delta_l \cr
~ 0	& ~ 0 ~	& -i\delta_l \cr
- i\delta_l	&  ~ i\delta_l ~	& 0 \cr } \right ) \; ;
%		(17)
\end{eqnarray}
and the non-hermitian perturbations
\begin{eqnarray}
\Delta {\bf M}^{(2)}_l & = & \frac{c^{~}_l}{3} \left ( \matrix{
-i\delta_l	& 0	& i\delta_l \cr
0	& i\delta	& -i\delta_l \cr
i\delta_l	& -i\delta_l	& 0 \cr } \right ) \; .
\nonumber \\
\Delta \tilde{\bf M}^{(2)}_l & = & \frac{c^{~}_l}{3} \left ( \matrix{
-i\delta_l	& ~ 0 ~	& ~~ 0 ~~ \cr
0	& ~ i\delta_l ~	& ~~ 0 ~~ \cr
0	& ~ 0 ~		& ~~ 0 ~~ \cr } \right ) \; ,
\nonumber \\
\Delta \hat{\bf M}^{(2)}_l & = & \frac{c^{~}_l}{3} \left ( \matrix{
~ 0 ~	& ~ 0	& i\delta_l \cr
~ 0 ~	& ~ 0	& -i\delta_l \cr
~ i\delta_l ~	& -i\delta_l	& 0 \cr } \right ) \; ,
%		(18)
\end{eqnarray}
The three hermitian and three non-hermitian perturbative
mass matrices obey the following sum rules: 
\begin{eqnarray}
\Delta {M}^{(2)}_l & = & \Delta \tilde{M}^{(2)}_l ~ + ~
\Delta \hat{M}^{(2)}_l \; , \nonumber \\
\Delta {\bf M}^{(2)}_l & = & \Delta \tilde{\bf M}^{(2)}_l
~ + ~ \Delta \hat{\bf M}^{(2)}_l \; .
%		(19)
\end{eqnarray}
Let us remark that hermitian perturbations of the same
forms as given in Eqs. (12) and (17) have been used to break the
flavor democracy of quark mass matrices and to generate $CP$
violation \cite{FP90,Lehmann}. The key point of this similarity
between the charged lepton and quark mass matrices is that both of
them have the strong mass hierarchy and might have the same
dynamical origin or a symmetry relationship.

To be more specific we transform all the six charged lepton mass matrices
\begin{eqnarray}
M^{(2)}_l & = & M^{(0)}_l + \Delta M^{(1)}_l + \Delta M^{(2)}_l \; ,
\nonumber \\
\tilde{M}^{(2)}_l & = & M^{(0)}_l + \Delta M^{(1)}_l + \Delta \tilde{M}^{(2)}_l \; ,
\nonumber \\
\hat{M}^{(2)}_l & = & M^{(0)}_l + \Delta M^{(1)}_l + \Delta \hat{M}^{(2)}_l \; 
%		(20)
\end{eqnarray}
and
\begin{eqnarray}
{\bf M}^{(2)}_l & = & M^{(0)}_l + \Delta M^{(1)}_l + \Delta {\bf M}^{(2)}_l \; ,
\nonumber \\
\tilde{\bf M}^{(2)}_l & = & M^{(0)}_l + \Delta M^{(1)}_l + 
\Delta \tilde{\bf M}^{(2)}_l \; ,
\nonumber \\
\hat{\bf M}^{(2)}_l & = & M^{(0)}_l + \Delta M^{(1)}_l + 
\Delta \hat{\bf M}^{(2)}_l \;
%		(21)
\end{eqnarray}
into their counterparts in the hierarchical basis
and list the results in Table 1.
A common feature of these hierarchical mass matrices is that
their (1,1) elements all vanish.
For this reason the $CP$-violating effects, resulted from
the hermitian perturbations $\Delta M^{(2)}_l$,
$\Delta \tilde{M}^{(2)}_l$, $\Delta \hat{M}^{(2)}_l$ and
the non-hermitian perturbations
$\Delta {\bf M}^{(2)}_l$, $\Delta \tilde{\bf M}^{(2)}_l$,
$\Delta \hat{\bf M}^{(2)}_l$ respectively, 
are approximately independent 
of other details of the flavor symmetry breaking and have
the identical strength to a high degree of accuracy.
Indeed it is easy to check that all the six charged lepton mass matrices in
Eqs. (20) and (21), together with the neutrino mass matrix
$M^{(2)}_\nu = M^{(0)}_\nu + \Delta M^{(1)}_\nu +
\Delta M^{(2)}_\nu$, lead to the same flavor mixing pattern $V$
as given in Eq. (15). Hence it is in practice
difficult to distinguish one scenario from
another. In our point of view, 
the similarity between $M^{(2)}_l$ 
and its quark counterpart \cite{FX99,Lehmann}
could provide us a useful hint towards an underlying
symmetry between quarks and charged leptons.
One may also argue that the simplicity of $\tilde{\bf M}^{(2)}_l$
and its parallelism with $M^{(2)}_\nu$ might make it technically
more natural to be derived from a yet unknown fundamental theory
of lepton mixing and $CP$ violation.

\small
%%%%%%%%%%%%%%%%%% Table %%%%%%%%%%%%%%%%%%%%
\begin{table}[t]
\caption{The counterparts of six charged lepton mass matrices 
in the hierarchical basis.}
\vspace{-0.5cm}
\begin{center}
\begin{tabular}{ccc} \\ \hline\hline 
%----------------------------------------------------------
$M^{(2)}_l/c^{~}_l$	& $\tilde{M}^{(2)}_l/c^{~}_l$ 
& $\hat{M}^{(2)}_l/c^{~}_l$ \\ \hline \\
%----------------------------------------------------------
$\left ( \matrix{
0	& -i\frac{1}{\sqrt{3}}\delta_l	& 0 \cr
i\frac{1}{\sqrt{3}}\delta_l	& \frac{2}{9}\varepsilon^{~}_l
& -\frac{\sqrt{2}}{9}\varepsilon^{~}_l \cr
0	& -\frac{\sqrt{2}}{9}\varepsilon^{~}_l
& 1 + \frac{1}{9}\varepsilon^{~}_l \cr} 
\right ) $ 
%-------------------
&
$\left ( \matrix{
0	& -i\frac{\sqrt{3}}{9}\delta_l	& -i\frac{\sqrt{6}}{9}\delta_l \cr
i\frac{\sqrt{3}}{9}\delta_l	& \frac{2}{9}\varepsilon^{~}_l
& -\frac{\sqrt{2}}{9}\varepsilon^{~}_l \cr
i\frac{\sqrt{6}}{9}\delta_l	& -\frac{\sqrt{2}}{9}\varepsilon^{~}_l
& 1 + \frac{1}{9}\varepsilon^{~}_l \cr} 
\right ) $
%----------------
& 
$\left ( \matrix{
0	& -i\frac{2\sqrt{3}}{9}\delta_l	& i \frac{\sqrt{6}}{9} \delta_l \cr
i\frac{2\sqrt{3}}{9}\delta_l	& \frac{2}{9}\varepsilon^{~}_l
& -\frac{\sqrt{2}}{9}\varepsilon^{~}_l \cr
-i \frac{\sqrt{6}}{9}\delta_l	& -\frac{\sqrt{2}}{9}\varepsilon^{~}_l
& 1 + \frac{1}{9}\varepsilon^{~}_l \cr} 
\right ) $ 
%-------------------------------------------------------
\\ \\ \hline\hline
${\bf M}^{(2)}_l/c^{~}_l$	& $\tilde{\bf M}^{(2)}_l/c^{~}_l$ 
& $\hat{\bf M}^{(2)}_l/c^{~}_l$ \\ \hline \\
%----------------------------------------------------------
$\left ( \matrix{
0	& -i\frac{1}{\sqrt{3}}\delta_l	& 0 \cr
-i\frac{1}{\sqrt{3}}\delta_l	& \frac{2}{9}\varepsilon^{~}_l
& -\frac{\sqrt{2}}{9}\varepsilon^{~}_l \cr
0	& -\frac{\sqrt{2}}{9}\varepsilon^{~}_l
& 1 + \frac{1}{9}\varepsilon^{~}_l \cr} 
\right ) $ 
%----------------
&
$\left ( \matrix{
0	& -i\frac{\sqrt{3}}{9}\delta_l	& -i\frac{\sqrt{6}}{9}\delta_l \cr
-i\frac{\sqrt{3}}{9}\delta_l	& \frac{2}{9}\varepsilon^{~}_l
& -\frac{\sqrt{2}}{9}\varepsilon^{~}_l \cr
-i\frac{\sqrt{6}}{9}\delta_l	& -\frac{\sqrt{2}}{9}\varepsilon^{~}_l
& 1 + \frac{1}{9}\varepsilon^{~}_l \cr} 
\right ) $
%----------------
& 
$\left ( \matrix{
0	& -i\frac{2\sqrt{3}}{9}\delta_l	& i \frac{\sqrt{6}}{9} \delta_l \cr
-i\frac{2\sqrt{3}}{9}\delta_l	& \frac{2}{9}\varepsilon^{~}_l
& -\frac{\sqrt{2}}{9}\varepsilon^{~}_l \cr
i \frac{\sqrt{6}}{9}\delta_l	& -\frac{\sqrt{2}}{9}\varepsilon^{~}_l
& 1 + \frac{1}{9}\varepsilon^{~}_l \cr} 
\right ) $
%-----------------------------------------------------------
\\ \\ \hline\hline
\end{tabular}
\end{center}
\end{table}
%%%%%%%%%%%%%%%%%%%%%%%%%%%%%%%%%%%%%%%%%%%%%%%%%%%%%%%%%%%%%%%%%%%%%%%%%%%%
\normalsize

The flavor mixing matrix $V$ can in general be parametrized
in terms of three Euler angles and one $CP$-violating phase
%%%%%%%%%%%%%%%%%%%%%%%%
\footnote{For neutrinos of the Majorana type, two additional
$CP$-violating phases may enter. But they are irrelevant 
to neutrino oscillations and can be neglected for our present
purpose.}. 
%%%%%%%%%%%%%%%%%%%%%%%%
A suitable parametrization reads as follows \cite{FX97}:
\begin{eqnarray}
V & = & \left ( \matrix{
c^{~}_l	& s^{~}_l	& 0 \cr
-s^{~}_l	& c^{~}_l	& 0 \cr
0	& 0	& 1 \cr } \right )  \left ( \matrix{
e^{-i\phi}	& 0	& 0 \cr
0	& c	& s \cr
0	& -s	& c \cr } \right )  \left ( \matrix{
c_{\nu}	& -s_{\nu}	& 0 \cr
s_{\nu}	& c_{\nu}	& 0 \cr
0	& 0	& 1 \cr } \right )  \nonumber \\ \nonumber \\
& = & \left ( \matrix{
s^{~}_l s_{\nu} c + c^{~}_l c_{\nu} e^{-i\phi} & ~~~~
s^{~}_l c_{\nu} c - c^{~}_l s_{\nu} e^{-i\phi} ~~~~ &
s^{~}_l s \cr
c^{~}_l s_{\nu} c - s^{~}_l c_{\nu} e^{-i\phi} &
c^{~}_l c_{\nu} c + s^{~}_l s_{\nu} e^{-i\phi}   &
c^{~}_l s \cr - s_{\nu} s	& - c^{~}_\nu s	& c \cr } \right ) \; ,
%		(22)
\end{eqnarray}
in which $s^{~}_l \equiv \sin\theta_l$, $s_{\nu} \equiv \sin\theta_{\nu}$, 
$c \equiv \cos\theta$, etc. The three mixing angles can all be
arranged to lie in the first quadrant, while the $CP$-violating
phase may take values between $0$ and $2\pi$.
It is straightforward to obtain
${\cal J} = s^{~}_l c^{~}_l s_\nu c_\nu s^2 c \sin\phi$. 
Numerically we find
\begin{equation}
\theta_l \approx 4^{\circ} \; , ~~~~~
\theta_\nu \approx 45^{\circ} \; , ~~~~~
\theta \approx 52^{\circ} \; , ~~~~~
\phi \approx 90^{\circ} \; .
%		(23)
\end{equation}
The smallness of $\theta_l$ is a natural consequence of the 
mass hierarchy in the charged lepton sector, just as the
smallness of $\theta_{\rm u}$ in quark mixing \cite{FX99}.
On the other hand, both $\theta_\nu$ and $\theta$ are too large
to be comparable with the corresponding quark mixing angles
(i.e., $\theta_{\rm d}$ and $\theta$ as defined in Ref. \cite{FX99}),
reflecting the qualitative difference between quark and lepton
flavor mixing phenomena. It is worth emphasizing that the
leptonic $CP$-violating phase $\phi$ takes a special value 
($\approx 90^{\circ}$) in our model. The same possibility
is also favored for the quark mixing phenomenon in a variety of
realistic mass matrices \cite{FX95}. 
Therefore maximal leptonic $CP$ violation, in the sense that
the magnitude of ${\cal J}$ is maximal for the fixed values
of three flavor mixing angles, appears naturally as in the
quark sector.

Some consequences of this lepton mixing scenario 
can be drawn as follows:

(1) The mixing pattern in Eq. (15), after neglecting small
corrections from the charged lepton masses, is quite similar to that
of the pseudoscalar mesons $\pi^0$, $\eta$ and $\eta'$ in QCD in
the limit of the chiral $SU(3)_{\rm L} \times SU(3)_{\rm R}$
symmetry \cite{FH94}:
\begin{eqnarray}
\pi^0 & = & \frac{1}{\sqrt{2}} \left (
|\bar{u}u\rangle - |\bar{d}d\rangle \right ) \; ,
\nonumber \\
\eta & = & \frac{1}{\sqrt{6}} \left (
|\bar{u}u\rangle + |\bar{d}d\rangle - 2 |\bar{s}s\rangle
\right ) \; ,
\nonumber \\
\eta' & = & \frac{1}{\sqrt{3}} \left (
|\bar{u}u\rangle + |\bar{d}d\rangle + |\bar{s}s\rangle 
\right ) \; .
%		(24)
\end{eqnarray}
A theoretical derivation of the flavor mixing matrix 
$V\approx U$ has been given in Ref. \cite{Mohapatra},
in the framework of a left-right symmetric extension of the
standard model with $S(3)$ and $Z(4) \times Z(3) \times Z(2)$
symmetries.

(2) The $V_{e3}$ element, of magnitude
\begin{equation}
|V_{e3}| \; =\; \frac{2}{\sqrt{6}} \sqrt{\frac{m_e}{m_\mu}} \;\; ,
%		(25)
\end{equation}
is naturally suppressed in
the symmetry breaking scheme outlined above.
A similar feature appears in the $3\times 3$ quark flavor mixing
matrix, i.e., $|V_{ub}|$ is the smallest among the
nine quark mixing elements. Indeed the smallness of $V_{e3}$
provides a necessary condition for the decoupling of
solar and atmospheric neutrino oscillations, even though neutrino
masses are nearly degenerate. The effect of small but nonvanishing
$V_{e3}$ will manifest itself in long-baseline $\nu_\mu
\rightarrow \nu_e$ and $\nu_e \rightarrow \nu_\tau$ transitions,
as already shown in Ref. \cite{FX96}.

(3) The flavor mixing between the 1st and 2nd lepton families
and that between the 2nd and 3rd lepton families are nearly
maximal. This property, together with the natural smallness
of $|V_{e3}|$, allows a satisfactory interpretation of the 
observed large mixing 
in atmospheric and solar neutrino oscillations. We obtain
%%%%%%%%%%%
\footnote{In calculating $\sin^2 2\theta_{\rm sun}$ we have
taken the ${\cal O}(m_e/m_\mu)$ correction to the 
expression of $V$ into account.}
%%%%%%%%%%
\begin{eqnarray}
\sin^2 2\theta_{\rm sun} & = & 1 - \frac{4}{3} \frac{m_e}{m_\mu} 
\; , \nonumber \\
\sin^2 2\theta_{\rm atm} & = & \frac{8}{9} +
\frac{8}{9} \frac{m_\mu}{m_\tau} \; 
%		(26)
\end{eqnarray}
to a high degree of accuracy. Explicitly
$\sin^2 2\theta_{\rm sun} \approx 0.99$ and $\sin^2 2\theta_{\rm atm}
\approx 0.94$, favored by current data \cite{SK}. It is obvious that the model
is fully consistent with the vacuum oscillation solution to
the solar neutrino problem \cite{Barger99} and might also be
able to incorporate the large-angle MSW 
solution \cite{Liu}.

(4) The rephasing-invariant strength of $CP$ violation in 
this scheme is given as
\begin{equation}
{\cal J} \; = \; \frac{1}{3\sqrt{3}} \sqrt{\frac{m_e}{m_\mu}}
\left ( 1 + \frac{1}{2} \frac{m_\mu}{m_\tau} \right ) \; .
%		(27)
\end{equation}
Explicitly we have ${\cal J}
\approx 1.4\%$. The large magnitude of $\cal J$ for lepton mixing
turns out to be very non-trivial, as the same quantity for quark mixing
is only of order $10^{-5}$ \cite{FX99,FX95}.  
If the mixing pattern under discussion 
were in no conflict with the large-angle MSW solution to the
solar neutrino problem, then
the $CP$- and $T$-violating signals 
$\Delta_{CP} = \Delta_T \propto -16 {\cal J} \approx -0.2$
could be significant enough to be measured from the asymmetry
between $P(\nu_\mu \rightarrow \nu_e)$ and 
$P(\bar{\nu}_\mu \rightarrow \bar{\nu}_e)$ or that
between $P(\nu_\mu \rightarrow \nu_e)$ and $P(\nu_e \rightarrow
\nu_\mu)$ in the long-baseline
neutrino experiments. In the leading-order approximation
we arrive at 
\begin{eqnarray}
{\cal A} & = & \frac{P(\nu_\mu \rightarrow \nu_e) ~ - ~ P(\bar{\nu}_\mu
\rightarrow \bar{\nu}_e)}{P(\nu_\mu \rightarrow \nu_e) ~ + ~
P(\bar{\nu}_\mu \rightarrow \bar{\nu}_e)} 
\nonumber \\ \nonumber \\
& = & \frac{P(\nu_\mu \rightarrow \nu_e) ~ - ~ P(\nu_e \rightarrow \nu_\mu)}
{P(\nu_\mu \rightarrow \nu_e) ~ + ~ P(\nu_e \rightarrow \nu_\mu)}
\nonumber \\ \nonumber \\
& = & \frac{\displaystyle -\frac{8}{\sqrt{3}} \sqrt{\frac{m_e}{m_\mu}}}
{\displaystyle \frac{16}{3} \frac{m_e}{m_\mu}
+ \left (\frac{\sin F_{12}}{\sin F_{23}} \right )^2} ~ \sin F_{12} \; .
%		(28)
\end{eqnarray}
The asymmetry ${\cal A}$ depends linearly on
the oscillating term $\sin F_{12}$, which is associated essentially with
the solar neutrino anomaly.

%%%%%%%%%%%%%%%%%%%% Fig. 1 %%%%%%%%%%%%%%%%
\begin{figure}
% GNUPLOT: LaTeX picture
\setlength{\unitlength}{0.240900pt}
\ifx\plotpoint\undefined\newsavebox{\plotpoint}\fi
\sbox{\plotpoint}{\rule[-0.200pt]{0.400pt}{0.400pt}}%

\vspace{0.4cm}
\caption{Illustrative plots for the $CP$-violating asymmetry 
$|\Delta_{CP}|$ 
(between $\nu_\mu \rightarrow \nu_e$ and $\bar{\nu}_\mu \rightarrow 
\bar{\nu}_e$ transitions) 
changing with the neutrino beam energy $E$, where
(a) $|\Delta m^2_{12}| =5\times 10^{-5} ~ {\rm eV}^2$ and (b) 
$|\Delta m^2_{12}| = 10^{-4} ~ {\rm eV}^2$ versus the fixed
$|\Delta m^2_{23}|
= 10^{-3} ~ {\rm eV}^2$ have typically been taken in the case
of the baseline length $L=732$ km or $L=7332$ km.}
\end{figure}
%%%%%%%%%%%%%%%%%%%%%%%%%%%%%%%%%%%%%%%%%%%%%%%%%%%%%%%%%%%%%%
To give one a numerical estimate of the magnitudes of  
$\Delta_{CP}$ and ${\cal A}$, we
typically take the baseline length to be $L = 732$ km or $L = 7332$ km
for a neutrino source at Fermilab pointing toward the Soudan mine 
in Minnesota or the Gran Sasso underground laboratory in 
Italy \cite{Muon}.
The mass-squared differences are chosen as
(a) $|\Delta m^2_{12}| = 5 \times 10^{-5} ~ {\rm eV}^2$ and (b)
$|\Delta m^2_{12}| = 10^{-4} ~ {\rm eV}^2$ versus the fixed
$|\Delta m^2_{23}| = 10^{-3} ~ {\rm eV}^2$. The behaviors of
the asymmetries $|\Delta_{CP}|$ (or $|\Delta_T|$) and
$|{\cal A}|$ changing with the beam energy $E$ in the range
$3 ~{\rm GeV} \leq E \leq 20 ~{\rm GeV}$ 
are then shown in Figs. 1 and 2, respectively. Clearly the asymmetry ${\cal A}$ 
can be of ${\cal O}(0.1)$ to ${\cal O}(1)$, even though the corresponding
magnitude of $\Delta_{CP}$ (or $\Delta_T$) is about two-order smaller.
In reality the matter effect on these $CP$ asymmetries should be
taken into account, in order to extract the genuine $CP$-odd
parameters. For the model under consideration,
the smallness of $|V_{e3}|$ ($\approx 0.056$) together with the maximal
$CP$ violating phase ($\phi \approx 90^{\circ}$) is expected to make 
the possible matter effect insignificant, and unable to completely
fake the genuine $CP$-violating signals \cite{Tanimoto99}.
%%%%%%%%%%%%%%%%%% Fig. 2 %%%%%%%%%%%%%%%%%%%%%%%%%%%
\begin{figure}
% GNUPLOT: LaTeX picture
\setlength{\unitlength}{0.240900pt}
\ifx\plotpoint\undefined\newsavebox{\plotpoint}\fi
\sbox{\plotpoint}{\rule[-0.200pt]{0.400pt}{0.400pt}}%

\vspace{0.4cm}
\caption{Illustrative plot for the $CP$-violating asymmetry $|{\cal A}|$
(between $\nu_\mu \rightarrow \nu_e$ and $\bar{\nu}_\mu \rightarrow
\bar{\nu}_e$ transitions) 
changing with the neutrino beam energy $E$, where (a) $|\Delta m^2_{12}|
= 5\times 10^{-5} ~ {\rm eV}^2$ and (b) $|\Delta m^2_{12}| =
10^{-4} ~ {\rm eV}^2$ versus the fixed 
$|\Delta m^2_{23}| = 10^{-3} ~ {\rm eV}^2$
have typically been taken in the case of the baseline length 
$L = 732$ km or $L = 7332$ km.}
\end{figure} 
%%%%%%%%%%%%%%%%%%%%%%%%%%%%%%%%%%%%%%%%%%%%%%%%%%%%%%%%%%%%%%%%%%

If the upcoming data appeared to rule out the consistency between our model
and the large-angle MSW solution to the solar neutrino problem, 
then it would be quite difficult
to test the model itself from its prediction for
large $CP$ and $T$ asymmetries in any realistic long-baseline
experiment. 

Finally it is worth remarking that our lepton mixing pattern has no conflict
with current constraints on the neutrinoless double beta 
decay \cite{Beta}, if neutrinos are of the Majorana type.
In the presence of $CP$ violation, the effective
mass term of the $(\beta\beta)_{0\nu}$ decay can be written as
\begin{equation}
\langle M\rangle_{(\beta\beta)_{0\nu}} \; = \;
\sum^3_{i=1} \left (m_i ~ \tilde{V}^2_{ei} \right ) \; ,
%		(31)
\end{equation}
where $\tilde{V} = VP_\nu$
and $P_\nu = {\rm Diag}\{1, e^{i\phi_1}, e^{i\phi_2} \}$ is the
Majorana phase matrix. If the unknown phases are taken to be 
$\phi_1 =\phi_2 =90^{\circ}$ for example, then one arrives at
\begin{equation}
\left | \langle M \rangle_{(\beta\beta)_{0\nu}} \right | \; = \; 
\frac{2}{\sqrt{3}} \sqrt{\frac{m_e}{m_\mu}}
~ m_i \; ,
%		(32)
\end{equation}
in which $m_i \sim 1 - 2$ eV (for $i=1,2,3$) as required by the 
near degeneracy of three neutrinos in our model
to accommodate the hot dark matter of the universe. 
Obviously $|\langle M\rangle_{(\beta\beta)_{0\nu}}| 
\approx 0.08 m_i \leq 0.2$ eV, the
latest bound of the $(\beta\beta)_{0\nu}$ decay \cite{Beta}.

In summary, we have extended our previous model of the
nearly bi-maximal lepton flavor mixing to incorporate large
$CP$ violation. The new model remains favored by current
data on atmospheric and solar neutrino oscillations, and
it predicts significant $CP$- and $T$-violating effects 
in the long-baseline neutrino experiments. We expect that
more data from the Super-Kamiokande and other neutrino
experiments could soon provide stringent tests of the
existing lepton mixing models and give useful hints 
towards the symmetry or dynamics of lepton mass generation.

\end{document}